# New stable crystal structures of C and GeC$_2$ predicted from first-principles calculations

Ying Yang, [1, 2*] Guang Yang, [1, 3] Xihong Peng[1*]

[1] College of Integrative Sciences and Arts, Arizona State University, Mesa, Arizona 85212, USA

[2] Department of Electronic Engineering, Xi'an University of Technology, Xi'an, Shaanxi, 710048, P. R. China

[3] Hebei University of Science and Technology, Shijiazhuang, 050018, P. R. China

## ABSTRACT

Two novel three-dimensional (3D) crystal structures of carbon (C) and germanium carbide (GeC$_2$) were predicted using first-principles density-functional theory (DFT) calculations. These newly discovered 3D carbon allotrope and GeC$_2$ are in the space group of *Fmmm* (space group number 69). Their crystal structures have unique tetragonal/hexagonal rings formed by either C or Ge/C atoms. Both structures were proven to be thermodynamically stable through the phonon spectrum calculations. The C allotrope is a semiconductor with a wide band gap of 3.65 eV predicted by the hybrid density functional HSE06 method, while GeC$_2$ is metallic. This indirect band gap of 3.65 eV in the C allotrope is between that in diamond and graphite and in the ultra-violet (UV) region. The new C allotrope has a low mass density of 2.84 g/cm$^3$ and is proven to be energetically stable with cohesive energy less than -7.5 eV/atom, which is lower than many other carbon allotropes. Such a carbon crystal structure with a low mass density and wide band gap once synthesized would have wide applications in gas adsorption sensors and photo-electronic devices.

*To whom correspondence should be addressed, yangy@xaut.edu.cn (Y. Y.); xihong.peng@asu.edu (X. P.)

In group IVA of the Periodic Table of the Element, carbon, silicon, and germanium are three important elements for semiconductor materials. In particular, carbon in $sp$, $sp^2$, and $sp^3$ hybridization, has much flexibility of bonding and exists in tremendous forms of allotropes [1-15], which exhibit numerous unique physical and chemical properties. These carbon allotropes can be classified into four different groups according to their band gap and mass density, namely (1) wide gap and high density, (2) narrow gap and high density, (3) narrow gap and low density, and (4) wide gap and low density. A typical representation in group (1) is the well-studied diamond with a band gap of 5.4 eV and a mass density of 3.54 g/cm$^3$ [1]. In addition to diamond, many other carbon allotropes in group (1) also have a band gap larger than 3 eV and a mass density higher than 3 g/cm$^3$, such as P carbon (3.64 eV and 3.41 g/cm$^3$) [2], h-diamond (3.13 eV and 3.52 g/cm$^3$) [3], M carbon (3.56 eV and 3.45 g/cm$^3$) [4], C10 (4.6 eV and 3.55 g/cm$^3$) [5], 8-tetra (2, 2) (3.76 eV and 3.3 g/cm$^3$) [6,7], PM1, PM2, and PM3 (3.07-3.09 eV and 3.38-3.48 g/cm$^3$) [8], D carbon (4.33 eV and 3.2 g/cm$^3$) [9], and protomene (3.38 eV and 3.5 g/cm$^3$) [10]. An example in group (2) allotrope having a narrow band gap and high density is novamene with a band gap of 0.3 eV and a mass density of 3.4 g/cm$^3$ [11]. Many additional allotropes were found in the group (3) with a narrow band gap and low density. For instance, CY carbon (1.93 eV and 2.27 g/cm$^3$) [12], TY carbon (1.5 eV and 0.52 g/cm$^3$) [13], tp8, tp12, tp16, tp24, op12, and op18 (1.48 - 2.9 eV and 2.07-2.5 g/cm$^3$) [14]. Lastly, the group (4) allotropes with a wide band gap larger than 3 eV and low density smaller than 2.2 g/cm$^3$, have the following examples, T carbon (3 eV and 1.5 g/cm$^3$) [15], Y carbon (4.6 eV and 0.89 g/cm$^3$) [12], and pentadiamond (3.31 eV and 2.2 g/cm$^3$) [1]. In general, the carbon structures with a high mass density also have a high hardness property [5], while the structures with a low mass density exhibit porous property [14]. If the carbon allotropes have a porous structure along with a wide band gap (i.e. the above-mentioned group (4) allotropes), they are expected to have potential applications in gas adsorption and photo-electronic devices. In this work, we proposed a new three-dimensional (3D) carbon crystal structure with a relatively low mass density and a wide band gap. It is proven to be thermodynamically stable and is energetically more favorable than most of the reported porous carbon allotropes in literature.

In addition to the classification of carbon allotropes according to their properties such as band gap and mass density, many literature work also assorts the structures according to their geometrical compositions, such as hexagonal, tetragonal, pentagonal, octagonal, heptagonal carbon rings, and some combinations of the mentioned rings. For example, the following structures show varying geometrical composition of carbon rings, 6-membered rings [3], 4+8 membered rings [2], 4+6+8 membered rings [8], 4+5 membered rings [9], and 5+7 membered rings [2], and so on. In this work, our proposed new 3D carbon structure consists of a unique tetragonal and hexagonal (4+6 membered) carbon rings (named as TH-C), which is constructed on an AB stacking of a recently discovered two-dimensional (2D) monolayer $sp^2$- and $sp^3$-hybridized carbon allotrope known as tetrahex-C [16]. Similarly, based on this tetrahex structure and AB stacking forms, we also propose a new 3D $GeC_2$ crystal structure with tetragonal and hexagonal rings named as 3D TH-$GeC_2$. This 3D $GeC_2$ is predicted to be metallic, opposite to the semiconducting nature of the multilayer honeycomb GeC crystal structure [17, 18].

The first-principles density-functional theory (DFT) [19] calculations were performed using the VASP code [20, 21]. The Perdew-Burke-Ernzerhof (PBE) exchange-correlation functional [22] was chosen for electronic structure calculations and geometry relaxation. Standard DFT calculations including PBE are well known for underestimating the electronic band gap of semiconductors. Heyd-Scuseria-Ernzerhof (HSE06) hybrid functional [23-25] was chosen for an improved prediction of the band gap. The projector-augmented wave (PAW) [26] potentials were used to treat interactions between ion cores and valance electrons. The reciprocal space is meshed using Monkhorst-Pack method [27]. The plane wave kinetic energy cutoff 900 eV and $15\times13\times14$ mesh for reciprocal space was chosen in geometry relaxation and force fields calculations along with the PBE functional. The energy convergence criterion for electronic iterations is set to be $10^{-8}$ eV and the force is converged within $10^{-4}$ eV/Å with the energy criterion $10^{-5}$ eV for geometry optimization of the unit and conventional cells. The kinetic energy cutoff 500 eV for plane wave basis set was used for the HSE06 band structure calculations.

The optimized crystal structures of the 3D TH-C and TH-$GeC_2$ are in the space group of *Fmmm* (space group number 69) and are displayed in Fig. 1. Fig. 1(a) (c) show the primitive cells of 3D TH-C and TH-$GeC_2$, respectively, which contain 6 atoms in the primitive cells. The parameters of the crystal structures are listed in Table 1. For an enhanced intuitive representation of the crystal structures, orthogonal conventional cells of both structures are also presented in Fig. 1(b) (d) with the optimized lattice constants *a, b*, *c* and bond angles $\theta_1$-$\theta_4$ listed in Table 1. For convenience to compare the 3D structures with their 2D monolayer and bilayer counterparts in tetrahex-C, we label the carbon atoms as C1 and C2 using brown and orange balls in the 3D TH-C in Fig. 1(a) (b). In the 2D counterparts of tetrahex-C, C1 and C2 represent $sp^2$- and $sp^3$-hybridized C atoms, respectively [16]. However, in the 3D TH-C structure, C1 atoms are no longer $sp^2$- hybridized. Instead, they form covalence bonds between interlayers and are $sp^3$-hybridized with a distorted networking.

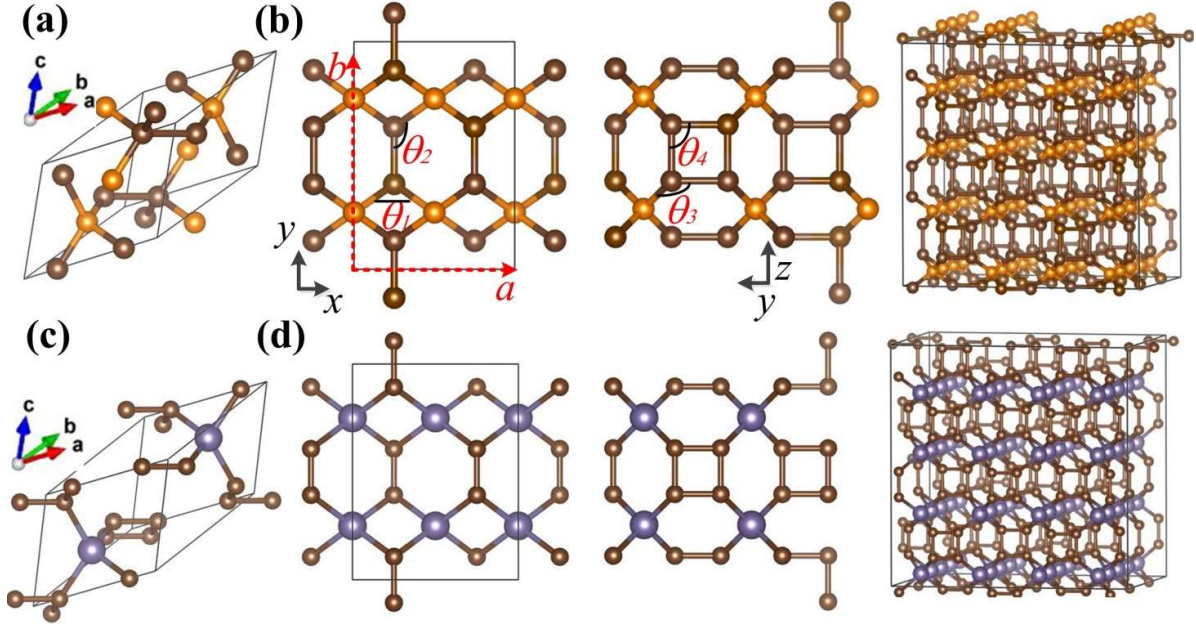

Fig.1. The optimized crystal structures of 3D TH-C and TH-GeC$_2$, (a) (c) the primitive cells of 3D TH-C and TH-GeC$_2$, respectively, (b) (d) the orthogonal conventional cells of 3D TH-C and TH-GeC$_2$, respectively, viewed in different orientations and in supercells. The blue, brown, and orange balls represent Ge, C1, and C2 atoms, respectively.

As shown in Fig. 1(b), there are two types of tetragonal (hexagonal) carbon rings in the 3D TH-C crystal structure. Type-I tetragonal (hexagonal) carbon rings lay in the $xy$-plane, which are similar to that in its 2D counterpart. Type-II tetragonal (hexagonal) carbon rings appear in the $yz$-plane. The bond angle $\theta_1$ denoted in Fig. 1(b) describes the shape of the type-I tetragonal rings and it was found that $\theta_1 = 90°$, which is smaller than that (95°) in its 2D tetrahex-C [16]. The bond angle $\theta_4$ depicting the type-II tetragonal rings in the $yz$-plane also has a value of 90°. On the other hand, for the 3D TH-GeC$_2$, the type-I tetragonal rings consist of 2 Ge and 2 C atoms, and the bond angle $\theta_1$ (86.9°) in the 3D TH-GeC$_2$ is 7.9° smaller than that (94.8°) in its 2D TH-GeC$_2$ counterpart [28], while the type-II tetragonal rings in the $yz$-plane consist of 4 C atoms with the bond angle $\theta_4$ still being 90°. The variation of $\theta_1$ in both 3D structures indicates that the type-I tetragonal rings in the $xy$-plane are more significantly distorted in the TH-GeC$_2$ than that in TH-C, which is expected to play a role in determining electronic properties of the 3D TH-GeC$_2$ structure.

The C-C bond length in the 3D TH-C has a value of 1.55 Å for both types of bonds, i.e. C1-C2 and C1-C1, which is almost as same as the C-C bond length in diamond [14]. However, in the 3D TH-GeC$_2$, it was found that the bond length Ge-C = 2.04 Å and C-C = 1.56 Å. Based on the optimized crystal structures, we can calculate the mass density of the structures. The 3D TH-C is predicted to have a mass density of 2.84 g/cm$^3$. It is almost in the middle of that in diamond (3.52 g/cm$^3$) and graphite (2.22 g/cm$^3$), implying that the 3D TH-C can be used as a porous material.

*Table 1. The optimized parameters of the primitive and conventional cells of the discovered C and GeC$_2$.*

| Parameters | Primitive Cell | | Conventional Cell | |
|---|---|---|---|---|
| | C | GeC$_2$ | C | GeC$_2$ |
| Space group | Fmmm-69 | Fmmm-69 | Pna21-33 | Pna21-33 |
| $a$ (Å) | 4.38 | 5.16 | 4.38 | 5.62 |
| $b$ (Å) | 3.80 | 4.61 | 6.20 | 7.30 |
| $c$ (Å) | 3.80 | 4.61 | 6.20 | 7.30 |
| $\alpha$ (°) | 70.5 | 68.2 | 90 | 90 |
| $\beta$ (°) | 54.8 | 55.9 | 90 | 90 |
| $\gamma$ (°) | 54.8 | 55.9 | 90 | 90 |
| $\rho$ (g/cm$^3$) | 2.84 | 4.28 | 2.84 | 4.28 |
| $\theta_1$ (°) | - | - | 90 | 86.9 |
| $\theta_2$ (°) | - | - | 120 | 120.9 |
| $\theta_3$ (°) | - | - | 120 | 120.9 |
| $\theta_4$ (°) | - | - | 90 | 90 |

Both 3D TH-C and TH-GeC$_2$ structures are proven to be thermodynamically stable via the calculations of their phonon spectra. Fig. 2 (a) and (b) present their phonon spectra and no imaginary phonon frequencies are observed indicating their dynamical stability. We also studied an analogical structure of SiC$_2$ and its phonon spectrum in Fig. 2(c) shows imaginary modes, suggesting this 3D SiC$_2$ structure is unstable.

To further investigate the energetically stability of the new 3D TH-C and TH-GeC$_2$ structures, we calculated their cohesive energy $E_{coh}$ using Eq. (1), and compared their values with existing diamond/germanium single crystal and other carbon allotropes in literature.

$$E_{coh} = \frac{E_{tot}(\text{Ge}_x\text{C}_y) - x\ E_{\text{Ge}} - y\ E_{\text{C}}}{x+y}, \qquad (1)$$

where $E_{tot}$, $E_{\text{Ge}}$, $E_{\text{C}}$ are the total energy of the material, and the energies of an isolated germanium and carbon atoms, respectively. Fig. 2(d) plots their cohesive energies. The dashed line connecting the energies of germanium single crystal and diamond is used to gauge the cohesive energy and an energy above the line implies a metastable structure. The new 3D TH-C and TH-GeC$_2$ are both metastable. The 3D TH-C has higher cohesive energy by 0.4 eV/atom compared to that in diamond. However, its cohesive energy is lower than most of reported carbon allotropes in literature. Its cohesive energy is lower by 0.4 eV, 0.69 eV, 0.75 eV, and 0.8 eV than that in CY carbon [12], Y carbon [12], TY carbon [13], and T carbon [15], respectively, suggesting the new 3D TH-C more energetically stable than those carbon allotropes..

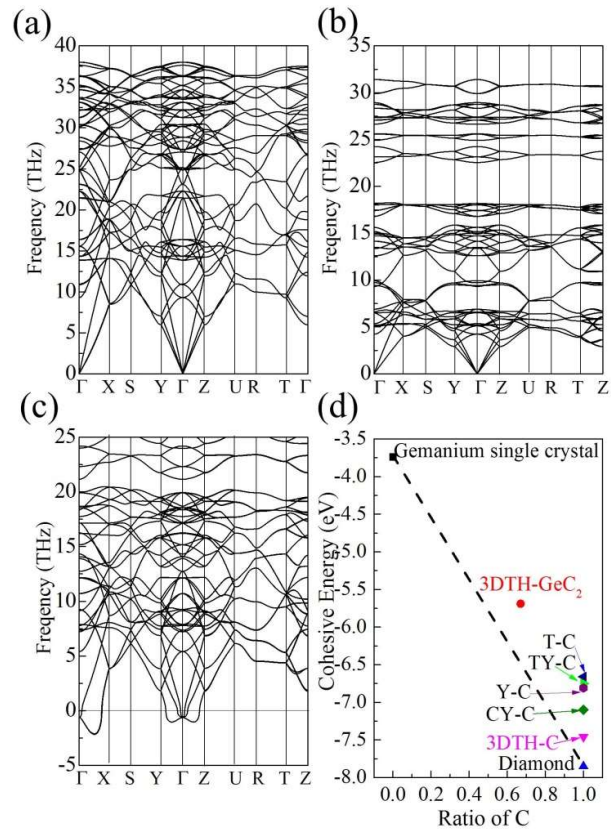

*Fig.2 (a)-(c) Phonon spectra and (d) cohesive energy of the predicted 3D structures. (a) 3D TH-C; (b) 3D TH-GeC$_2$; (c) 3D TH-SiC$_2$; (d) their cohesive energies compared to other carbon allotropes in literature. The dashed line connecting the energies of germanium single crystal and diamond is a guide for eye. For structural details, one can refer to pertinent papers (T-C [15], TY-C [13], Y-C [12], and CY-C [12]).*

Fig. 3 presents the electronic band structures of the 3D TH-C and GeC$_2$. The 3D TH-C possesses an indirect band gap of 3.65 eV predicted by HSE06 functional in Fig. 3(a), whereas GeC$_2$ shows metallic property (HSE06) in Fig. 3 (c). The band gap in the 3D carbon is determined by the energy difference of the valence band maximum (VBM, i.e. state A as denoted in Fig. 3(a)) and the conduction band minimum (CBM, i.e. state B). The projected band structure in Fig.3 (b) demonstrates that the state A (VBM) in 3D TH-C is mainly contributed by the $p_x$, $p_z$ and $p_y$ orbitals of C1 and C2 atoms. State A is a three-fold-degeneracy state at the T point. However, along the crystal direction of T-Γ as shown in Fig. 3(b), the state of $p_x$ orbitals contributed from C1 atoms (green lines) are separated from another two-fold-degeneracy states which correspond to the $p_z$ and $p_y$ orbitals (degenerated and overlapped) from the C1 and C2 atoms, respectively. Whereas, state B (CBM) in 3D TH-C is mainly contributed by the $p_x$ and $s$ orbitals of C1 atoms. The band structure of the 3D TH-GeC$_2$ in Fig. 3(c) display crossings of conduction/valence bands at the Fermi level indicating a metallic characteristic. Its projected band structure in Fig. 3(d) illustrates that the state A is mainly contributed by the $p_x$ orbitals of C/Ge atoms, while the state B is dominated by the $p_x$ orbital of C atoms and the $s$ orbitals of Ge atoms. Both states A and B are nondegenerate. From Fig. 3, we speculated that the metallic nature of GeC$_2$ is resulted from the aforementioned structural distortion which leads to the upward shift of the energy of State A to exceed the Fermi level and become a half-full band.

To verify the above speculation, the electron density contour plots of state A in the carbon and GeC$_2$ crystals are presented in Fig. 4. For an enhanced intuitive visualization and convenient comparison with their 2D counterparts, the orthogonal conventional cells of the 3D C and GeC$_2$ were used to demonstrate the electron density. As mentioned previously, the state A (VBM) in the 3D TH-C are three-fold degenerate and the electron cloud distribution of the three modes are displayed in Fig. 4(a)-(c), respectively. One has electron density mainly located in the bonds C1-C2 as shown in Fig. 4(a), the other two are distributed around the bonds of C1-C1 atoms (Fig. 4(b) and (c)). These two types of electron density distributions are related to two kinds of σ bonds in the 3D TH-C, respectively. They correspond to the two different bands along the direction of T-Γ in Fig. 3(b). The electron density distribution between C1-C2 bonds in Fig. 4(a) corresponds to the green band in Fig. 3(b), i.e. the $p_x$ orbitals of C1 and C2 atoms. The electron density distributions between C1-C1 bonds in Fig. 4 (b) (c) correspond to the overlapped and degenerated blue and red bands in Fig. 3(b). Fig. 4(b) and (c) also reveal that the electron density distribution between C1-C1 bonds can be subdivided into two categories with dominated $p_y$ and $p_z$ orbitals as shown in Fig. 4 (b) and (c), respectively. These two types of electron density distribution correspond exactly to the energy bands contributed by the $p_y$ (i.e. the red band in Fig. 3(b)) and $p_z$ (i. e. the blue band in Fig. 3(b)) orbitals of C atoms, respectively.

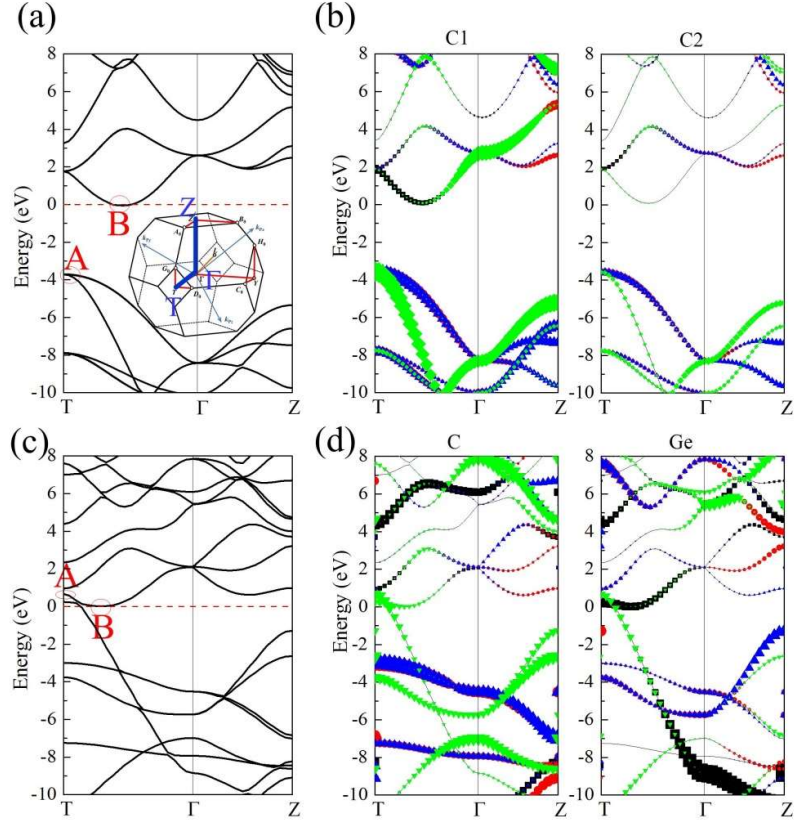

Fig. 3 (a) (c) The electronic band structures of the 3D TH-C and TH-GeC$_2$, respectively. (b) The projected band structure of C1 (left) and C2 atoms (right) in the 3D TH-C. (d) The projected band structure of C (left) and Ge atoms (right) in the GeC$_2$. The black, red, blue, and green lines in (b) (d) represent the s, $p_y$, $p_z$, and $p_x$ orbitals of each element, respectively. Fermi level is set at zero. The inset in (a) shows the Brillion zone of the primitive cell.

The electron density of state A in the 3D TH-C (Fig. 4(a)) has a similar distribution as that of state A in the relaxed 2D tetrahex-C [29]. However, there is a big energy difference between these two states. The state A in the 2D tetrahex-C structure is related to the pure covalent σ bonds and it is buried deep inside of the valence bands due to stability [30]. However, in the 3D TH-C structure, the state A appears on the top of the valence band representing the VBM. The higher energy of state A in the 3D TH-C is resulted by the intrinsic strain [29] of structural distortion presented in the structure.

The band gap in the 2D monolayer of tetrahex-C is predicted to be 2.63 eV (direct gap) [16] and the bilayer of tetrahex-C possesses a bigger gap of 3.1 eV [16]. And in this 3D TH-C, the band gap increases to 3.65 eV and is indirect. Such an increase of the band gap is due to the formation of covalent bonding between the interlayer along in the z direction. The VBM and CBM in the 2D and bilayer tetrahex-C structures are contributed by the $p_z$ orbitals of the $sp^2$-hybridized C atoms [16, 30]. It means that the band gap is determined by the energy difference between π and π* bonds [30]. In general, π bonds are less stable than σ bonds. Therefore, the energy difference between π and π* bonds is relatively smaller than that between σ and σ* bonds. The VBM/CBM has more contributions from the π/π* bonds than σ/σ* bonds, implying a narrower band gap. Consequently, with the increase of the layer numbers of tetrahex-C, the contributions from the π/π* bonds vanish gradually, and the band gap becomes wider. In the 3D C structure, all the π/π* bonds are removed through forming the interlayer covalent bonds. Therefore, the band gap reaches up to 3.65 eV. The formation of the interlayer covalent bonding also indicates all of the $sp^2$-hybridized C atoms in the 2D structures are fully transferred into $sp^3$-hybridizzation, which eventually leads to the modification in the band edge states shifted from Γ (2D) to Γ-T (3D) due to the interactions between in-plane as well as out-of-plane orbitals.

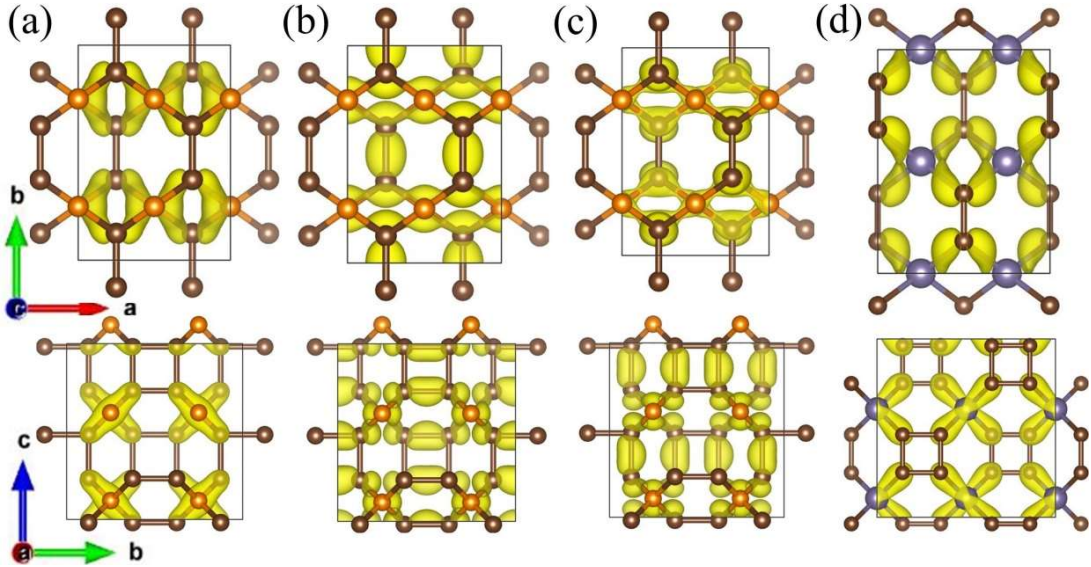

Fig. 4 (a)-(c) The electron density contour plots of state A with its three-fold-degeneracy modes in 3D TH-C. The isosurface was set to be 0.005 e/Å³. (d) The electron density contour plot of the state A in the 3D TH-GeC$_2$. The isosurface was set to be 0.003 e/Å³. Top rows are in the xy-plane and bottom rows in the yz-plane.

Fig. 4(d) presents the electron density contour plot of state A at T in the 3D TH-GeC$_2$ structure. It was found that the 2D counterpart of TH-GeC$_2$ is a semiconductor with a direct band gap of 0.84 eV [28]. And the state A shown in Fig. 4(d) in the 3D structure has a similar charge distribution as that of VBM in its 2D counterpart [28] which implies that state A in 3D TH-GeC$_2$ is related to the VBM of its 2D structure. According to Fig. 4(d), state A in 3D TH-GeC$_2$ is mainly contributed by the σ bonds between Ge and C atoms with a relatively high energy, which are coordination σ bonds with originally high energy. In addition, as mentioned previously, the bond angle $θ_1$ in the 3D TH-GeC$_2$ is reduced by 7.9° compared to its 2D structure [28]. This leads to a structural distortion and intrinsic strain presented in the 3D structure, which further increases the energy of state A to exceed the Fermi level to become a half-full band. Therefore, the 3D TH-GeC$_2$ shows a metallic behavior.

In summary, new 3D crystal structures of carbon and GeC$_2$ are proposed via first principles DFT calculations. The structures are proven to be thermodynamically stable and energetically more favorable compared to many other well-known 3D carbon allotropes such as T-carbon, TY-carbon, Y-carbon, and CY-carbon in literature indicating a possibility of synthesis in labs. The 3D carbon is a semiconductor with an indirect band gap of 3.65 eV and GeC$_2$ is metallic. The band gap in the 3D carbon is smaller than that in diamond but wider than that in graphite. The evolution of the electronic band structures including the band gap in the 3D C/GeC$_2$ and their 2D counterparts were investigated in detail. The new materials consist of tetragonal and hexagon rings pure C or Ge/C atoms. The unique 4+6 membered-ring carbon structure has a mass density of 2.84 g/cm$^3$, which is almost in the middle of the mass densities in diamond and graphite. Low mass density in the structure indicates that it may be used as a porous semiconducting material, which has wide applications in photo-electronic devices, gas sensors, *etc*. It is of importance to explore possible synthesis methods and further study other properties of this material, such as mechanical stiffness, charge effective masses and mobility, high-temperature stability.

**Acknowledgement**

This work is financially supported by the National Natural Science Foundation of China (Grant No.51207128). The authors thank Arizona State University Advanced Computing Center for providing computing resources (Agave Cluster).

The authors declare no competing financial interest. Data available on request from the authors.